
%
%
%
%
%
%
\def\figcond{1}     
\ifnum\figcond>0 
\input epsf
\fi
\normalbaselineskip=12pt
\baselineskip=12pt
\magnification=1200
\overfullrule=0pt
\hsize 6.0true in
\vsize 8.5true in
\parskip=0pt plus 2pt
\font\lbf = cmbx10 scaled \magstep 1
\def\lsim{\mathrel{\rlap{\lower4pt\hbox{\hskip1pt$\sim$}}
    \raise1pt\hbox{$<$}}}         
\def\gsim{\mathrel{\rlap{\lower4pt\hbox{\hskip1pt$\sim$}}
    \raise1pt\hbox{$>$}}}         

\def\overleftrightarrow#1{\vbox{\ialign{##\crcr
    $\leftrightarrow$\crcr
    \noalign{\kern 1pt\nointerlineskip}
    $\hfil\displaystyle{#1}\hfil$\crcr}}}
\long\def\caption#1#2{{\setbox1=\hbox{#1\quad}\hbox{\copy1%
\vtop{\advance\hsize by -\wd1 \noindent #2}}}}
\line{\hfill DOE/ER/40427-08-N94}
\vskip 15pt
\centerline{{\lbf MESONS AND THE STRUCTURE OF NUCLEONS}\footnote*{It is
an honor for us to dedicate this article to the memory of Bob
Marshak, who was a direct inspiration to one of us and a mentor to
many young physicists.
\medskip
\noindent
$^\dagger$Supported in part by the Department of Energy, Grant No.
DE-FG06-88ER40427.
\medskip
\noindent
$^{\ddagger}$Supported in part by the Deutsche Forschungsgemeinschaft
and by the Feodor-Lynen program of the Alexander von Humboldt-Stiftung.
\medskip
\noindent
$^{\dagger\dagger}$Supported in part by the National Science
Foundation, Grant No. PHY-9223618.}}
\vskip 15pt
\centerline{W. KOEPF$^{(1,   \dagger,\ddagger)}$, 
         E.M. HENLEY$^{(1,2, \dagger)}$,
       and M. ALBERG$^{(1,3, \dagger\dagger)}$}
\vskip 15pt
\centerline{\it $^{(1)}$Department of Physics, FM-15, University of
Washington}
\centerline{\it  Seattle, Washington  98195 USA}
\vskip 9pt
\centerline{\it $^{(2)}$Institute for Nuclear Theory, HN-12,
University of Washington}
\centerline{\it Seattle, Washington  98195 USA}
\vskip 9pt
\centerline{\it $^{(3)}$Department of Physics, Seattle University,
Seattle, WA 98122 USA}
\vskip 24pt
\centerline{ABSTRACT}
{\narrower
\smallskip
The role of mesons, particularly the pion, in the structure of
nucleons is reviewed and investigated.  Since quark-antiquark pairs
are likely to ``transform" into mesons at large distances, mesons are
expected to contribute to nucleon structure.  Their effects on the
Gottfried sum rule, on the strangeness content of the nucleon, and on
the spin of the nucleon are discussed.}
\baselineskip=14pt

\bigskip
\noindent
{\lbf 1.  Introduction}
\medskip

Ever since the postulation of mesons by Yukawa in 1934, and the
discovery of the pion in 1947, it has been clear that mesons play a
crucial role in the structure of the nucleon and in hadronic forces.
The long range part of the nucleon-nucleon 
interaction is clearly dominated by
meson exchanges.  The puzzle of the short-distance behavior of
nuclear forces has been replaced by our lack of knowledge of the
transition from quarks and gluons to nucleons and mesons.  Although
it is now accepted that QCD is the basis of hadronic forces, the
translation of that knowledge to calculations of hadronic properties
is beset by difficulties.  Thus, models of the structure of nucleons
abound.  Many of these models make use of constituent quarks, which
represent quarks ``dressed" with gluons and quark-antiquark pairs.
Since mesons are composed of these pairs, it is natural to believe
that at large distances ($\approx$ 1 fm) the quark-antiquark pairs 
become mesons.  Direct evidence for the role of the pion in nucleon
structure comes, for instance, from the charge form factor of the
neutron, $G_E^n$, at small momentum transfers.$^1$ The negative
charge of the pion $(n \to p \pi^-)$ is responsible for the long
range part of the charge distribution and accounts for the
experimentally observed negative charge radius of the neutron.

Further evidence for the role of mesons in nucleon structure includes
a) the success of models in which they play an explicit role, such as
the cloudy bag model;$^1$ and b) the experimental value for the
Gottfried sum rule.$^2$  In addition, the EMC-inspired ``spin
crisis" led to the proposition that strange quarks contribute
non-negligibly to the spin of the nucleon and that only a small
fraction of the nucleon's spin is due to the valence quarks.$^{3,4}$
These conjectures brought about suggestions for further experiments to
measure the strangeness ``content" of the nucleon$^{5-7}$ and the
corresponding unknown form factors;$^7$ they also led to calculations
of the meson contribution to the strangeness of the nucleon,$^8$ and
to more recent measurements in deep inelastic scattering on $^3$He
and deuterons.$^9$

In this paper we intend to review and expand on some of this work.

\bigskip
\noindent
{\lbf 2. Mesons and the Gottfried Sum Rule}
\medskip

The Gottfried sum rule relates to the $F_2$ structure functions of
the proton and neutron$^{10}$, 
$$S_G = \int^1_0 dx \, {[F_2^p (x) - F_2^n (x)] \over x} \ , \eqno(1)$$ 
where $F_2$ is determined from deep inelastic charged ($\mu$ or $e$)
lepton-nucleon scattering.  In the quark-parton model, $S_G$ can be
written as 
$$S_G = \sum_i \left({e_i \over e}\right)^2 \int^1_0 \, \left[q^p_i
(x) + \bar q_i^p (x) - q^n_i (x) - \bar q_i^n (x) \right] dx 
\ , \eqno(2)$$ 
where $e_i/e$ is the charge of quark $i$ in units of the elementary
charge. Since
charge symmetry has been shown to be valid for hadrons to $\lsim
1/2\%$, we accept it here; it follows that $d^n = u^p \equiv u, \,
\bar d^n = \bar u^p \equiv \bar u, \, u^n = d^p \equiv d$, and $\bar 
u^n = \bar d^p \equiv \bar d$.  However charge symmetry does not imply
flavor independence of the sea, i.e., $\bar u = \bar d$.  With these
simplifications, Eq.  (2) becomes 
$$
S_G \ = \ {1 \over 3} + {2 \over 3} \int^1_0 [\bar u (x) - \bar
d (x)]dx \ = \ {1 \over 3} (1 - \Delta\bar q) \ , \eqno(3)$$
where the last integral is the sea quark contribution.  The
experimental result of the NMC group,$^2$ $S_G \approx 0.240 \pm
0.016$ at $Q^2$ = 5 GeV$^2$, leads to the conclusion that there
are more $\bar d$ quarks than $\bar u$ quarks in the sea of the
proton $(\Delta\bar q = 0.280 \pm 0.048)$, unless charge symmetry
fails.  This result appears at first sight to be a puzzle, since
gluons would be expected to generate an equal number of $u \bar u$
and $d \bar d$ pairs if quark mass difference effects are neglected.
However, as first pointed out by Field and Feynman,$^{11}$ the Pauli
exclusion principle, alone, would give an excess of $d \bar d$ pairs
over $u \bar u$ pairs in the proton sea since there are already
two valence
$u$ quarks and only one $d$ quark in the proton.  However, it is
doubtful that this effect would lead to as large a reduction of $S_G$
as observed.  On the other hand, the pion cloud contribution also
affects $S_G$.  Although the $\pi^0$ has an equal number of $\bar u$
and $\bar d$ quarks, the proton has an excess of $\pi^+ (u \bar d)$
over $\pi^- (d \bar u)$ and thus more $\bar d$ quarks than $\bar u$
quarks.$^{12}$  Of course, when the $\pi^+$ is ``in the air", the
valence quark distribution of the proton is altered; however, for
small $x$, the distinction between valence and sea quarks is
difficult to maintain.  On the constituent quark level, the above
process is $u \to d + ( u \bar d).$ Since there are more $u$ quarks
than $d$ quarks $(d \to u + d \bar u)$, the same asymmetry holds at
the constituent quark level.

\midinsert
\vbox to 2.55truein{
\ifnum\figcond>0 
\centerline{\hbox{\hsize=6truein\hss\vbox to 2.45truein{\vss
\epsfxsize=6.0truein
\vskip -0.45truein
\centerline{\epsfbox{f1.ps}}
\vss}\hss}}
\else
\centerline{\hbox{\hsize=5truein\vrule\hss\vbox to 2.45truein{\hrule\vss
\hss
\vss\hrule}\hss\vrule}}
\fi
\vfill}
\baselineskip=12pt
\narrower{\noindent
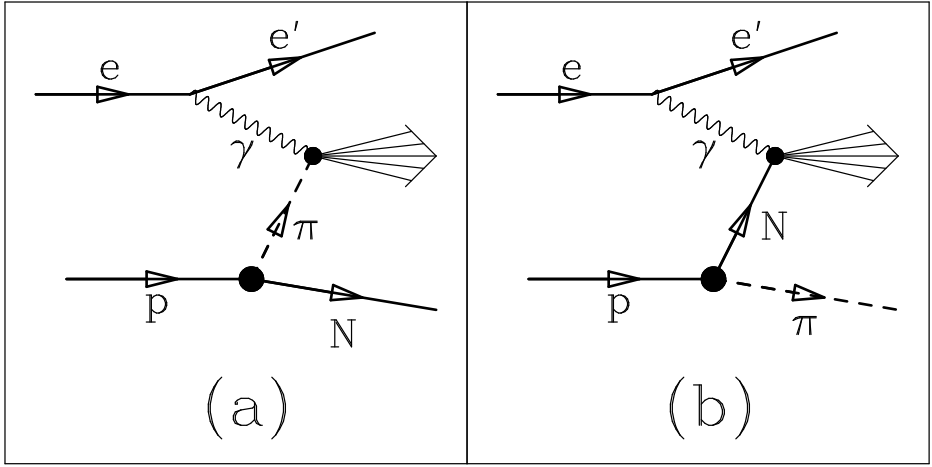
{\bf Figure 1.} The pion cloud contribution to the proton structure 
function in deep inelastic scattering from the 
$p \rightarrow N + \pi$ process.}
\endinsert
\baselineskip=14pt

\midinsert
\vbox to 2.55truein{
\ifnum\figcond>0 
\centerline{\hbox{\hsize=6truein\hss\vbox to 2.45truein{\vss
\epsfxsize=6.0truein
\vskip -0.45truein
\centerline{\epsfbox{f2.ps}}
\vss}\hss}}
\else
\centerline{\hbox{\hsize=5truein\vrule\hss\vbox to 2.45truein{\hrule\vss
\hss
\vss\hrule}\hss\vrule}}
\fi
\vfill}
\baselineskip=12pt
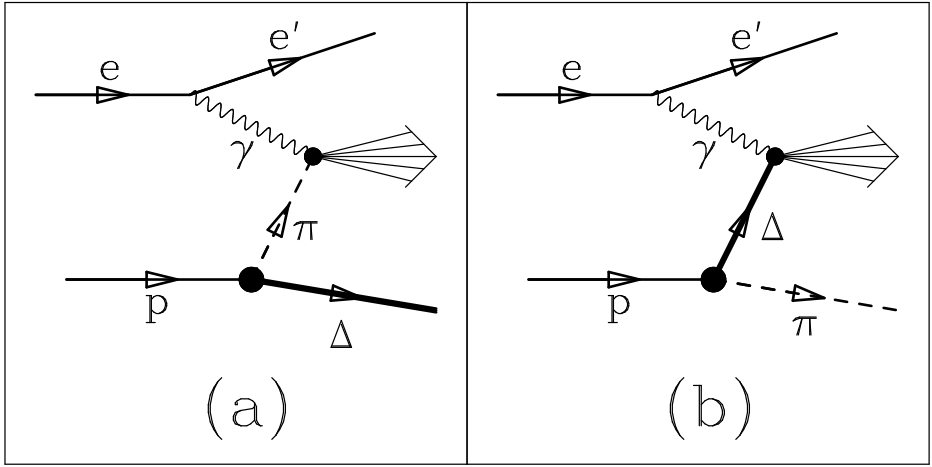
\centerline{{\bf Figure 2.} Same as Fig. 1, but from the $p \rightarrow 
\Delta + \pi$ process.}
\endinsert
\baselineskip=14pt

The meson cloud contribution to the structure function was considered
by Sullivan$^{13}$ over 20 years ago and has received considerable
revived interest in the past few years.$^{12,14}$  The pionic
contribution, Figs. 1 and 2, to the proton structure function in deep
inelastic scattering can be written as a convolution$^{12,14}$ of the
pion structure function, $F^\pi_2$, and its momentum distribution,
$f_\pi$, in the infinite momentum frame.  We find,
$$\eqalignno{\Delta F^p_2 (x)& = \int^1_x \, dy \, f_\pi (y) F_2^\pi
\left({x \over y},Q^2\right) + \int_0^{1-x} \, dy \, f_\pi (y) \, F^N_2 
\, \left({x \over 1-y},Q^2 \right), &(4) \cr
f_\pi (y) & = {3 g^2_{\pi N} \over (4 \pi)^2} \, \int^\infty_{{M^2 y 
\over 1 - y}} \, {t \over (t + m^2_\pi)} \, F^2_{\pi N} (t) \, dt 
\,, &(5) }$$
\noindent 
where $g_{\pi N}$ is the pion-nucleon coupling constant, $M$ the
nucleon mass, and $F_{\pi N}$ is the pion-nucleon form factor.  The
first term in Eq. (4) comes from Fig.  1a and the second one from
Fig. 1b, where the photon couples to the recoil nucleon.  In our
approach, when the difference between a proton and neutron is taken,
the symmetry of a photon striking the $\pi^+$ in the proton and the
$\pi^-$ in the neutron cancels this contribution to the Gottfried sum
rule and leaves only the contribution of the second term in Eq. (4),
shown in Fig.~1b.

In addition to the contribution of Fig. 1 to the structure function,
pion emission from a nucleon can lead to excited states.
The most important of these is likely to be that of lowest energy,
the $\Delta$, where $p \to \Delta^{++} \pi^-, \Delta^+ \pi^0$,
$\Delta^0 \pi^+$ and $n \to \Delta^- \pi^+, \Delta ^0 \pi^0$.  Here,
the contribution of the pion cloud, shown in Figs. 2a and 2b,
tends to increase $S_G$ above 1/3.
The contribution to $\Delta F_2$ is similar to Eqs. (4) and (5) with
$N$ replaced by $\Delta$.  The structure function of the $\Delta$ is
less well known than that of the nucleon and the pion and thus there is more
uncertainty in calculating this contribution.  However, it is expected
to be smaller due to the mass difference of the nucleon and delta
states which increases the minimum value of $t$ in Eq. (5) to
$[M^2_\Delta - (1 - y) M^2 ] y /(1-y)$. We neglect higher mass
resonances and other mesons. If we include the pion contribution, 
the dressed nucleon state function $|N \rangle$ is 
$$|N \rangle = \sqrt{{\cal N}} \left\{|N) + \sqrt{n_\pi} | N \pi) +
\sqrt{\Delta_\pi} | \Delta \pi) \right\} , \eqno(6)$$ 
where $|N)$ is a bare nucleon.  The structure function $F_2$ is then
renormalized by the factor ${\cal N} = \left(1 + n_\pi +
\Delta_\pi\right)^{-1}$, with $n_\pi = \int^1_0 \, dy \, f_\pi (y) \,
dy$ and similarly for $\Delta_\pi$.  The technical details are
described in Refs.  12-14.  With these approximations, the Gottfried
sum rule becomes 
$$S_G = {\cal N} \left({1 \over 3} - {n_\pi \over 9} 
+ {5 \Delta \pi \over 9} \right) \,. \eqno(7)$$

The numerical value of $S_G$ depends on the form factor $F_{\pi N}$;
the higher the effective momentum cutoff, the larger the reduction
of $S_G$ from 1/3; the experimental value of 0.24 can be reached for
a cutoff of $\approx$ 1.5 GeV.

In our opinion, the two mechanisms, the Pauli principle and the pion
cloud, offer natural explanations for the measured reduction of the
Gottfried sum rule from the anticipated value of 1/3.

Recent experiments have attempted to measure the pion contribution to
the excess of $\bar d$ (over $\bar u$) quarks in a Drell-Yan
reaction.$^{15}$  Although no asymmetry from the pions was observed,
this conclusion may not be inconsistent with the results of the NMC
group and our interpretation thereof.$^{16}$

\bigskip
\noindent
{\lbf 3.  Mesons and Strangeness in the Nucleon}
\medskip

There has been considerable interest in the strangeness matrix
elements in the nucleon, sparked, on the one hand, by a determination
of the strange quark scalar density, $\langle N|\bar s s|N\rangle$, from the
pion-nucleon sigma term.  Although the interpretation of the
experimental results has been the source of much debate,$^{17}$ it
points at a sizeable value of $\langle N|\bar s s|N\rangle$, which, in turn,
may be the reason behind violations of the OZI rule observed in the
$\bar p p \rightarrow \Phi+X$ reaction.$^{18}$  Knowledge of the
matrix element $\langle N|\bar s s|N\rangle$ would also be of importance
in determining the critical density for kaon condensation in dense
nuclear matter and hence for the occurrence of enhanced strange
particle production in heavy ion collisions.$^{19}$

In addition, there exists experimental evidence for a non-vanishing
strange quark axial vector matrix element, $\langle N|\bar s \gamma_\mu
\gamma_5 s|N\rangle$, through the BNL low-energy elastic neutrino-proton
cross sections$^{20}$ as well as from the EMC measurements$^3$ of the
spin dependent structure function of the proton in deep inelastic
scattering of polarized muons on polarized hydrogen.  The latter
determination of the strange axial vector form factor relies on
$SU(3)$ symmetry arguments, and it has been re-examined through
measurements carried out recently by the SMC$^{21}$ and the
E142$^{22}$ groups on deep inelastic scattering of polarized leptons
on a polarized deuteron and $^3$He target, respectively.  See Refs. 9
and 23 for clear discussions on this topic.

Furthermore, the goal of the SAMPLE experiment presently underway at
MIT-Bates$^{24}$ as well as of three other experiments employing
parity violating electron scattering on either protons or $^4$He
planned for CEBAF$^{25}$ is  to constrain the strange quark vector
matrix element, $\langle N|\bar s \gamma_\mu  s|N\rangle$.  Also, a new
determination of the strange quark axial vector form factor at a
significantly lower momentum transfer than in the original BNL
measurements is expected from the LSND experiment at LAMPF,$^{26}$
and there are further suggestions for even more exotic measurements
of the various strange quark matrix elements in the nucleon.$^5$

From a theoretical standpoint, the existence of significant strange
quark matrix elements in the nucleon is rather surprising, especially
in the light of the success with which naive constituent quark models
-- which inherently disregard the existence of any ``strangeness" in
the nucleon -- account for most low-energy properties of the baryonic
octet.  In a recent article, Karl pointed out that there
is actually no contradiction between the baryon magnetic moments and
the existence of non-vanishing strangeness matrix elements in the
nucleon, even in a constituent quark picture, if one is willing to
accept that the constituent quarks themselves have a non-trivial
structure.$^{27}$ Steininger and Weise showed that such a picture arises
quite naturally in a chiral quark model,$^{28}$ and Kaplan and
Manohar pointed out that a different multiplicative renormalization
in different flavor $SU(3)$ representations as generated by the
$U_A(1)$ axial anomaly can lead to a non-trivial flavor structure of
the constituent quarks, and hence to strange quark matrix elements in
non-strange hadrons.$^4$

Aside from those more qualitative than quantitative investigations,
there exists, however, only a handful of theoretical calculations
that pre\-sent an estimate of the strangeness matrix elements of the
nucleon.  There is Jaffe's pole analysis$^{29}$ where the
strangeness radius and the strange quark magnetic moment were obtained
using three-pole dispersion theory fits to the nucleon's isoscalar
form factor, together with a standard treatment of the $\Phi$-$\omega$
mixing and some mild assumptions on the asymptotic behavior of the
nucleon form factors.  Similar ideas were utilized in a calculation
by Cohen et al.,$^{30}$ where the authors consider the strangeness
vector current in the  nucleon as generated through the $\Phi$-$\omega$
mixing in a model based on  vector meson dominance (VMD). In
addition, there are both Skyrme$^{31}$ as well as kaon loop
model$^{8,32,33}$ estimates of the various strange quark matrix
elements in the nucleon.

In the following, we will concentrate on the amount of intrinsic
strange\-ness which is generated in the nucleon through its virtual
kaon cloud, i.e., by means of the dissociation of a nucleon into a
strange baryon plus a kaon, as e.g., by the process
$$p~\rightarrow~\Lambda~+~K^+~\rightarrow~p \ . \eqno(8)$$

In such a picture, one should not at all be surprised that strange
quark matrix elements are non-vanishing, since there exist
substantial $\hat J_s = \bar s \Gamma s$ couplings (with $\Gamma =
1$, $\gamma_\mu$ or $\gamma_\mu \gamma_5$) to the kaon as well as to
the strange hyperons of the baryonic octet and decuplet.  In Refs.
8, 32, and 33, the nucleon's strange quark matrix elements which arise
from its virtual kaon cloud were studied in one-loop calculations,
and each time quite different prescriptions for the evaluation of the
relevant loop integrals were employed.

In the first part of Ref. 8, a single meson-loop model was studied in
covariant perturbation theory, and pseudoscalar point couplings
between free nucleon and meson fields were used.  The meson cloud
contribution to both the usual electromagnetic as well as to the
strange magnetic moment was calculated, since the corresponding form
factor, $F_2$, is explicitly finite even without any cutoff.  The
model, which treats the bare hadrons as pointlike structureless
particles, leads to a significant strangeness content ($F_2^s(0)
\approx -1.0~\mu_N$) but it gives very unsatisfactory results for
the magnetic moments, especially in the isoscalar channel.  We find
$\kappa^{IS} \equiv \kappa^p + \kappa^n = -2.4$, which is an order of
magnitude larger than  the experimental value of $-0.12$.

The kaon loops in this model contribute at a
sizeable level, because the relevant integrals are dominated by the
baryonic masses; yet the model fails to provide a quantitative
description of the standard electromagnetic properties of the
nucleon.  This arises, in part, because the model does not properly
take into account the underlying nucleon structure and its spatial
extension.  When the quark sub-structure is considered, it is the
size of the proton, rather than its mass, that determines the
effective momentum cutoff.

In the second part of Ref. 8, the authors thus study models which
include such a cutoff -- as set by confinement -- and which have been
shown to give a very good description of the static properties of the
nucleon, namely an $SU(3)$ extended cloudy bag model (CBM)$^{34}$ as
well as a chiral non-relativistic constituent quark model (CNRQM).$^{35}$
The only free parameter in the model -- the MIT bag radius for
the CBM and the oscillator length for the chiral non-relativistic
constituent quark model -- was adjusted to give an optimal
description of the standard electromagnetic observables, i.e.  the
charge radii and magnetic moments of both the neutron and proton.
Predictions were then made for the strange quark contributions to the
weak magnetic form factor, $F_2^s(0)$, the strangeness radius,
$\langle r^2 \rangle_s$, and the strangeness axial vector matrix-element,
$g_A^s$.  The corresponding results, including some ``seagull" gauge
invariance corrections which were not considered in the original work
in Ref. 8, will be presented at the end of this section for a MIT bag
radius of $R \approx 1.1$ fm adjusted to fit the electromagnetic
observables and which, in turn, corresponds to an effective
three-momentum cutoff of approximately 0.2 GeV.  For further details
see Ref. 8 and the references therein.

In the meson-loop calculation of Musolf and Burkardt,$^{32}$ on the
other hand, no attempt was made to reproduce the electromagnetic form
factors.  Rather, the authors employ an effective meson-nucleon
vertex characterized by a monopole form factor
$$F(k^2)~=~{m^2 - \Lambda^2 \over k^2 - \Lambda^2} \ , \eqno(9)$$
where the ultraviolet cutoff parameter $\Lambda$ is obtained from
nucleon-nucleon and nucleon-hyperon scattering, and is typically in
the range of 1 to 2 GeV.  In a similar evaluation of the kaon loop
contribution, Holstein$^8$ equates the UV cutoff parameter with the
nucleon mass, and he finds a moderate value for the strange axial
vector form factor of $g_A^s \approx -0.1$.

\medskip
\baselineskip=12pt
\midinsert
\narrower{\noindent
{\bf Table 1.} Estimates for the strange quark vector and
axial vector matrix elements in the nucleon, deduced both from
experiment (the first two rows) as well as from different hadronic
models.  The last four rows correspond to the ``virtual kaon loop"
calculations. Results from Ref. 32 were obtained with a cutoff
of $\Lambda \approx$ 1.3 GeV.}
$$
\offinterlineskip \tabskip=0pt
\vbox{
\halign to 0.90 \hsize
   {\strut
   \tabskip=0pt plus 15pt               %
   # \hfil                              
 & \vrule#                              
 & \ \hfil # \hfil                      
 & \hfil # \hfil                        
 & \hfil # \hfil                        
 & \vrule#                              
 & \ \hfil # \hfil                      
   \tabskip=0pt                         %
   \cr                                  %
\noalign{\hrule}                        
&&&&&&\cr
&&& 
\hbox to 1pt{\hss$\langle N|\bar s \gamma_\mu s|N \rangle$\hss}
&&&
$\langle N|\bar s \gamma_\mu \gamma_5 s |N\rangle$ \cr
&&&&&&\cr
&& $F^s_2 (0)$ $\left[\mu_N\right]$ && $\langle r^2 \rangle_s$ [fm$^2$] 
&& $g^s_A$ \cr
&&&&&&\cr
\noalign{\hrule}                 
&&&&&&\cr
BNL$^{20}$ &&&&&& $-0.15 \pm 0.09$ \cr
EMC$^3$ &&&&&& $-0.19 \pm 0.06$ \cr
&&&&&&\cr
Poles$^{29}$ && $-0.31 \pm 0.09$  && $0.14 \pm 0.07$ &&\cr
VMD$^{30}$   &&&& $-0.04$ &&\cr
Skyrme$^{31}$ && $-0.33 \dots -0.13$ && $-0.21 \dots -0.11$ && $-0.10$ \cr
&&&&&&\cr
Point Couplings$^8$ && $-1.0$ &&&&\cr
Monopole Cutoff$^{32}$ && $-0.3$ && $-0.03$ && $-0.04$ \cr
Cloudy Bag$^8$ && $-0.09$ && $-0.006$ && $-0.004$ \cr
CCDM$^{33}$ && $-0.03$ && $-0.004$ && $-0.009$ \cr
&&&&&&\cr
\noalign{\hrule}                
}}$$
\endinsert
\baselineskip=14pt

In Table 1, we summarize the various estimates for the strange quark
form factors of the nucleon deduced both from experiments as well as
from some of the theoretical model calculations discussed above.
Recently, Phatak and Sahu$^{33}$ evaluated the nucleonic strangeness
content in a $SU(3)$ chiral color dielectric model (CCDM), which is
in spirit very similar to the extended cloudy bag model which was
studied in Ref. 8.  Note, however, that both in
Ref. 33 as well as in the original work in Ref. 8, the Ward-Takahashi
identity was not satisfied, and that the corresponding ``seagull"
corrections have an appreciable effect on the matrix elements under
consideration.$^{8,32}$  For further details on the issue of gauge
invariance in chiral hybrid quark models see Refs.  8 and 32, and
references therein.

As expected, the ``virtual kaon loop" estimates are rather sensitive
to the momentum cutoff used in the effective meson-nucleon vertices,
and they diminish rapidly with decreasing cutoff parameter $\Lambda$.
Except for the calculation employing point couplings for these
vertices, which, in turn, leads to quite unsatisfactory predictions
for the standard electromagnetic observables, the strange quark
matrix elements that were obtained in the loop analyses are
significantly smaller than the ones found either in Jaffe's pole
analysis$^{29}$ or in the $SU(3)$ Skyrme model calculations.$^{31}$

This is not at all surprising, since in most models discussed here
the mesonic loops yield only small corrections to hadronic models
that themselves were designed to already fit the bulk of the
low-energy properties of the baryonic octet.  Due to the relatively
high mass of the kaon, when compared e.g., with a natural energy
scale set by the size of the nucleon or the mass of the pion, its
contribution -- and hence the amount of strangeness in the nucleon
generated by means of virtual kaon loops -- will be rather small.

We are well aware that the loop calculations discussed here can only
yield a crude estimate for the strange quark matrix elements under
consideration.  Not only do they attempt to use a perturbative
expansion for intrinsically strongly coupled physics, but they also
neglect the contribution from the $U_A(1)$ axial anomaly and still
employ the traditional picture of a nucleon built from three
non-strange valence quarks, and thus discard any intrinsic
strangeness in the constituent quarks themselves, as suggested e.g.,
by Karl$^{27}$ or by Kaplan and Manohar.$^4$  However, we believe
that they point to evidence for significant meson contributions to
the nucleon structure.

\bigskip
\noindent
{\lbf 4.  Mesons and the Spin of the Nucleon}
\medskip
In the previous two sections we have discussed how the meson cloud
contributes to a reduction of $S_G$ in the Gottfried sum rule and to
the existence of non-zero strange quark matrix elements in the
nucleon.  The cloud also affects the spin of the nucleon, since the
dressed nucleon state (Eq.~(6)) includes terms in which mesons carry
orbital angular momentum and thereby depolarize the nucleon.  Mesons
play an important role in understanding the alleged ``spin crisis"
created by the EMC measurements$^{3}$ and further studied in the
SMC$^{21}$ and E142$^{22}$ experiments.  
In the following,
we review the experimental determination of the nucleon spin and the
theoretical interpretation of these results.

In the EMC experiment, measurements of deep inelastic scattering of
polarized muons from polarized hydrogen targets were used to
determine the asymmetry
$$A = {\sigma^\uparrow - \sigma^\downarrow \over \sigma^\uparrow +
\sigma^\downarrow} \ ,
\eqno(10)$$
in which $\sigma^{\uparrow (\downarrow)}$ is the cross section for
muons polarized parallel (antiparallel) to the spin of the proton.
From the asymmetry $A (x, Q^2)$ and the unpolarized structure
function $F_1 (x, Q^2)$ one can determine the polarized structure
function $g_1 (x, Q^2)$, where $x$ is the Bjorken scaling variable,
through
$$g_1 (x, Q^2) = F_1 (x, Q^2) \, A(x, Q^2) \ . \eqno(11)$$
In the limit of asymptotic momentum transfer, the polarized structure
function $g_1 (x)$ is determined by the distribution functions
$q^{\uparrow (\downarrow)}_i (x)$ of a quark $i$ with spin parallel
(antiparallel) to that of the proton,
$$g_1 (x) = {1 \over 2} \sum_i \, e^2_i \left[q_i^\uparrow \, (x) -
q^\downarrow_i (x)
\right] \ . \eqno(12)$$

The spin of the proton can be decomposed into 3 terms,
$${1 \over 2} = {1 \over 2} \sum_i \, \Delta q_i + \Delta G + 
\langle L_z \rangle \ , \eqno(13)$$
which correspond to the quark, gluon and orbital angular momentum
contributions, respectively.  The quark contributions 
$$\Delta q_i = \int^1_0 \, dx
\left(q^\uparrow_i + \bar q^\uparrow_i - q^\downarrow_i -
\bar q^\downarrow_i \right) + O (\alpha_s/\pi) \eqno(14)$$
are related to the first moments $\Gamma_1$ of the experimentally determined
polarized structure functions. In the asymptotic limit, we have e.g.,
$$\Gamma_1^p=\int^1_0 g^p_1 (x) dx = {1 \over 2} \left({4 \over 9} 
\Delta u + {1 \over 9} \Delta d + {1 \over 9} \Delta s \right) \ .
\eqno(15)$$
These first moments can be used, together with weak coupling
constants extracted from neutron and hyperon decays, to determine
the individual contributions $\Delta u$,$\Delta d$, and $\Delta s$.

The EMC experiment found
$$\Gamma_1^p = 0.126 \pm 0.011 \pm 0.014 \ , \eqno(16)$$
whereas the Ellis-Jaffe sum rule,$^{36}$ which assumes $\Delta s = 0$, gives
$$\Gamma_1^p = 0.175 \pm 0.007 \ . \eqno(17)$$
This was the origin of the ``spin crisis".  The experimental result
indicated that almost none of the proton's spin was carried by its
valence quarks, and that $\Delta s$ was significantly non-zero.
Intense theoretical and experimental activity was stimulated.  In
particular, a measurement of the neutron structure function $g^n_1$
was sought, since the most fundamental sum rule, derived by
Bjorken$^{37}$, relates the difference of the first moments of the
proton and neutron structure functions to the isovector axial vector
coupling constant,
$$\Gamma_1^p-\Gamma_1^n=\int^1_0 \left[g^p_1 (x, Q^2) - g^n_1 (x, Q^2) 
\right] dx = {g_A\over 6} \left(1 - {{\alpha_s (Q^2) \over\pi}+
O(\alpha_s^2) } \right) \ . \eqno(18)$$

The SMC experiment (deuteron target) and E142 experiment ($^3$He
target) were designed to provide data on the neutron structure
functions.  Their results appeared to contradict one another, with
SMC agreeing qualitatively with the EMC measurements, but E142
finding that the quarks carried about $1/2$ of the nucleon spin and
$\Delta s$ consistent with zero.  However, the EMC, SMC, and E142
experiments were carried out at different values of $Q^2$,
and since the $\Gamma_1^{p,n}$ are $Q^2$ dependent, care
must be taken to evolve data to the same
value of $Q^2$ before the first moments are taken.  Close and
Roberts,$^{23}$ Ellis and Karliner$^{38}$ and SMC$^{39}$ have
reanalyzed all the existing data.  They take into account improved
determinations of $F_1 (x, Q^2)$ from the NMC group, errors
introduced by extrapolations to low and high values of $x$, and
they include leading order QCD corrections.  They conclude that the
combined experimental data is consistent with the Bjorken sum rule.
Their analyses also show that the fraction of the nucleon spin carried by
quarks is $Q^2$ dependent, decreasing with increasing $Q^2$, and that
the strange quark contribution is significantly non-zero.

\midinsert
\vbox to 5.60truein{
\ifnum\figcond>0 
\centerline{\hbox{\hsize=6truein\hss\vbox to 5.5truein{\vss
\epsfxsize=6.0truein
\vskip -0.5truein
\centerline{\hskip 1.7truein \epsfbox{f3.ps}}
\vss}\hss}}
\else
\centerline{\hbox{\hsize=5truein\vrule\hss\vbox to 5.5truein{\hrule\vss
\hskip 5truein
\vss\hrule}\hss\vrule}}
\fi
\vfill}
\baselineskip=12pt
\narrower{\noindent
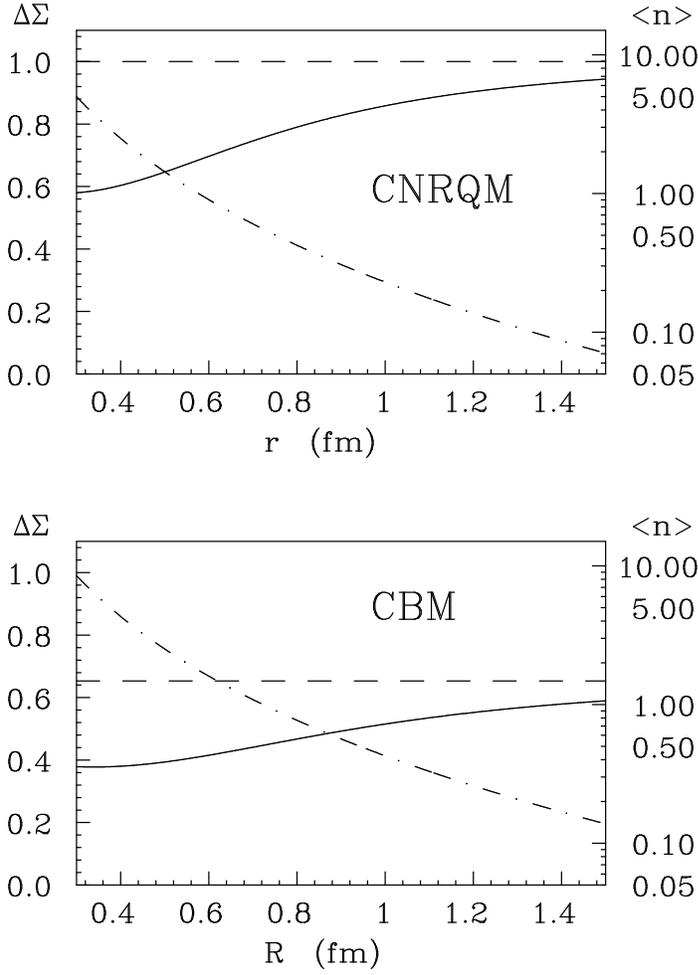
{\bf Figure 3.} The share of the nucleon's spin carried by its
valence quarks (solid line, left scale) and the average number of
``mesons in the air" (dotdashed line, right scale) as a function of
the oscillator length $r$ for the chiral non-relativistic constituent
quark model (upper figure) or the MIT bag radius $R$ for the cloudy
bag model (lower figure).  The dashed lines indicate the spin
contribution of the valence quarks when the depolarizing effects of
the meson cloud are neglected.}
\endinsert
\baselineskip=14pt

If the quark contribution to the nucleon's spin,
$$\Delta \Sigma \equiv \sum_i \, \Delta q_i ,\eqno(19)$$
in Eq.~(13) is small, then the other terms, $\Delta G$ and $\langle
L_z \rangle$, must compensate.  Contributions to $\langle L_z
\rangle$ are generated by relativistic effects and the meson cloud.
Confinement of the quarks in the nucleon leads to non-zero lower
components of the spinor wave functions, which carry angular momentum
even for quarks in a $s_{1/2}$ state, and the $\bar q q$ pairs in the
meson cloud also carry orbital angular momentum and thus depolarize
the nucleon.  These effects have been evaluated in the model
discussed in  Ref.~8.

In the non-relativistic quark model, $\Delta \Sigma$, the fraction of
the nucleon's spin carried by the valence quarks, is equal to one. In
the MIT bag model, $\Delta\Sigma = 0.65$ through the aforementioned
relativistic effects. These values are changed in
the one-meson-loop calculations of Ref.~8 through couplings to the
baryonic octet and decuplet and simultaneous
emission of a virtual pion or kaon.
Coupling to the octet reduces the axial current of the nucleon,
whereas coupling to the decuplet enhances it. The leading order
contribution is that of the octet, so the overall effect is a
reduction of $\Delta \Sigma$ that depends on the spatial extent of
the nucleon, characterized e.g., by the bag radius or oscillator length.
In Fig.~3, we show predictions for $\Delta \Sigma$ for the extended
cloudy bag model of Ref.~34 and for the chiral non-relativistic 
constituent quark
model of Ref.~35. As expected, $\Delta \Sigma$ is decreased by the
one-loop corrections, and the size of the effect decreases with
increasing nucleon size. The reduction is about 15\% for nucleon size
parameters (MIT bag radius $R\approx 1.1$ fm or oscillator length
$r\approx 0.8$ fm) which give a good description of the standard
electromagnetic observables. The dashed lines show the valence
quarks' contribution to the nucleon spin without the depolarization
effect of the mesonic cloud.  In this figure, we also depict the
average number of "mesons in the air",
$$\langle n \rangle = n_\pi + n_K + \Delta_\pi + \Delta_K \ , \eqno(20)$$ 
defined in accordance with Eq. (6).

Mesons thus contribute important corrections to our qualitative
understanding of the nucleon structure. Here, we have shown, in
particular, the effects of the virtual meson cloud on the reduction of
the Gottfried sum rule, the appearance of non-zero strange quark
matrix elements in the nucleon, and why the valence quarks'
contribution to the nucleon spin may be smaller than naively
anticipated.  Therefore, any complete, quantitative evaluation of
the nucleon structure should include the effects of the meson cloud.

\bigskip
\noindent
{\lbf 5. References}
\medskip
\baselineskip=12pt
\parindent=15pt

\item{1.} See e.g., G.A. Miller, {\it Intern. Rev. Nucl. Phys.} {\bf 1} 
(1984) 189 ; A.W. Thomas, {\it Prog. Nucl. Phys.} {\bf 13} (1984) 1.
\smallskip
\item{2.}P. Amaudruz et al., {\it Phys. Rev. Lett.} {\bf 66} (1991) 2712.
\smallskip
\item{3.}J. Ashman et al., {\it Phys. Lett.} {\bf B206} (1988) 364; {\it 
Nucl.  Phys.} {\bf B328} (1989) 1.
\smallskip
\item{4.}See e.g., D.B. Kaplan and A.V. Manohar, {\it Nucl. Phys.} {\bf 
B310} (1988) 527; R.L. Jaffe and A.V. Manohar, {\it Nucl. Phys} {\bf 
B337} (1990) 509; D.B. Kaplan, {\it Phys. Lett.} {\bf B275} (1992) 137.
\smallskip
\item{5.}W-Y.P. Hwang and E.M. Henley, {\it Ann. Phys. (NY)} {\bf 129} 
(1980) 47; E.M. Henley, G. Krein and A.G. Williams, {\it Phys. Lett.} 
{\bf B281} (1992) 178; T. Frederico et al., {\it Phys. Rev. } {\bf C46} 
(1992) 347; E.M. Henley, G. Krein, S.J. Pollock, and A.G. Williams, {\it 
Phys. Lett.} {\bf B269} (1991) 31.
\smallskip
\item{6.}G.T. Garvey et al., {\it Phys. Lett.} {\bf B289} (1992) 249; 
T. Suzuki, Y. Kohyama, and K. Yazaki, {\it Phys. Lett.} {\bf B252} 
(1990) 323.
\smallskip
\item{7.}D.H. Beck, {\it Phys. Rev.} {\bf D39} (1989) 3248; R. D. 
McKeown, {\it Phys. Lett.} {\bf B219} (1989) 140; E.J. Beise and R.D. 
McKeown, {\it Comm. Nucl. Part. Phys.} {\bf 20} (1991) 105.
\smallskip 
\item{8.}W. Koepf, E.M. Henley and S.J. Pollock, {\it Phys. Lett.} {\bf 
B288} (1992) 11; W. Koepf and E.M. Henley in print by {\it Phys. Rev.} 
{\bf C}; B.R. Holstein, {\it Proc. Parity Violation in Electron 
Scattering} eds. E.J. Beise and R.D. McKeown (World Scientific Singapore, 
1990) pp. 27-43.
\smallskip 
\item{9.}For an analysis and references, see G. Altarelli, P. Nason 
and G.  Ridolfi, {\it Phys. Lett.} {\bf B320} (1994) 152.
\smallskip
\item{10.}K. Gottfried, {\it Phys. Rev. Lett.} {\bf 18} (1967) 1174.
\smallskip
\item{11.}R.D. Field and R.P. Feynman, {\it Phys. Rev.} {\bf D15} (1977) 
2590, and {\it Nucl. Phys.} {\bf B136} (1978) 1.
\smallskip
\item{12.}A.W. Thomas, {\it Phys. Lett.} {\bf 126B} (1983) 97; 
E.M.Henley and G.A. Miller, {\it Phys. Lett.} {\bf B251} (1990) 453.
\smallskip
\item{13.}J.D. Sullivan, {\it Phys. Rev.} {\bf D5} (1972) 1732.
\smallskip
\item{14.}A.I. Signal and A.W. Thomas, {\it Phys. Lett.} {\bf B211} 
(1988) 481, and {\it Phys. Rev.} {\bf D40} (1989) 2832; A.I. Signal, 
A.W. Schreiber, and A.W. Thomas, {\it Mod. Phys. Lett.} {\bf A6} (1991) 
271, S. Kumano, {\it Phys. Rev.} {\bf D43} (1991) 59; S. Forte, {\it Phys. 
Rev.} {\bf D47} (1993) 1842; W. Melnitchouk, A.W. Thomas, and A.I. Signal, 
{\it Z. Phys.} {\bf A340} (1991) 85; S. Kumano and J.T. Londergan, {\it 
Phys. Rev.} {\bf D46} (1992) 457.
\smallskip
\item{15.}P.L. McGaughey et al., {\it Phys. Rev. Lett.} {\bf 69} (1992) 1726.
\smallskip
\item{16.}E.J. Eichten, I. Hinchliffe and C. Quigg, {\it Phys. Rev.} 
{\bf D47} (1993) R747;
A. Szczurek and J. Speth, {\it Nucl. Phys.} {\bf A555} (1993) 249.
\smallskip
\item{17.}T.P. Cheng and R.F. Dashen, {\it Phys. Rev. Lett.} {\bf 26} 
(1971) 594 and {\it Phys. Rev.} {\bf D13} (1976) 216; T.P. Cheng, {\it 
Phys. Rev.} {\bf D13} (1976) 2161; C.A. Dominguez and P. Langacker, {\it 
Phys. Rev.} {\bf D24} (1981) 190; J.F. Donoghue and C.R. Nappi, {\it 
Phys. Lett.} {\bf 168B} (1986) 105; J.F. Donoghue, {\it Ann. Rev. Nucl. 
Part. Sci.} {\bf 39} (1989) 1; J. Gasser, H. Leutwyler and M.E. Sainio, 
{\it  Phys. Lett.} {\bf B253} (1991) 252.
\smallskip
\item{18.} H. Genz and G. H\"ohler, {\it Phys. Lett.} {\bf 61B} (1976) 
389; J. Ellis, E. Gabathuler and M. Karliner, {\it Phys. Lett.} {\bf B217} 
(1989) 173; B.L. Ioffe and M. Karliner, {\it Phys. Lett.} {\bf B247} (1990) 
387; ASTERIX Collaboration, {\it Phys. Lett.} {\bf B267} (1991) 299.
\smallskip
\item{19.} D.B. Kaplan and A.E. Nelson, {\it Phys. Lett.} {\bf B175} 
(1986) 57; {\it ibid.} {\bf B179} (1986) 409(E); {\it ibid.} {\bf B192} 
(1987) 193.
\smallskip
\item{20.}L.A. Ahrens et al., {\it Phys. Rev.} {\bf D35} (1987) 785.
\smallskip
\item{21.} B. Adeva et al., {\it Phys. Lett.} {\bf B302} (1993) 533.
\smallskip
\item{22.} P.L. Anthony et al., {\it Phys. Rev. Lett.} {\bf 71} (1993) 959.
\smallskip
\item{23.} F.E. Close and R.G. Roberts, {\it Phys. Lett.} {\bf B316} 
(1993) 165.
\smallskip
\item{24.} Bates experiment No. 89-06, R.D. McKeown and D.H. Beck
spokespersons. 
\smallskip
\item{25.} CEBAF proposal No. PR-91-004, E.J. Beise spokesperson;
CEBAF proposal No. PR-91-010, J.M. Finn and P.A. Souder spokespersons;
CEBAF proposal No. PR-91-017, D.H. Beck spokesperson.
\smallskip
\item{26.} LAMPF proposal No. 1173, W.C. Louis, spokesperson.
\smallskip
\item{27.} G. Karl, {\it Phys. Rev.} {\bf D45} (1992) 247.
\smallskip
\item{28.} K. Steininger and W. Weise, {\it Phys. Rev.} {\bf D48} (1993) 1433.
\smallskip
\item{29.} R.L. Jaffe, {\it Phys. Lett.} {\bf B229} (1989) 275.
\smallskip
\item{30.} T.D. Cohen, H. Forkel and M. Nielsen, {\it Phys. Lett.} {\bf B316}
(1993) 1.
\smallskip
\item{31.} N.W. Park, J. Schechter and H. Weigel, {\it Phys. Rev.} {\bf D43}
(1991) 869.
\smallskip
\item{32.} M.J. Musolf and M. Burkardt, CEBAF preprint TH-93-01 (1993), in
print by {\it Z. Phys.} {\bf C}.
\smallskip
\item{33.} S.C. Phatak and S. Sahu, {\it  Phys. Lett.} {\bf B321} (1994) 11.
\smallskip
\item{34.} P. Zenczykowski, {\it Phys. Rev.} {\bf D29} (1984) 577 .
\smallskip
\item{35.} Y. Nogami and N. Ohtsuka, {\it Phys. Rev.} {\bf D26} (1982) 261.
\smallskip
\item{36.}J.Ellis and R.L. Jaffe, {\it Phys. Rev.} {\bf D9} (1974) 1444; 
{\it ibid.} {\bf D10} (1974) 1669.
\smallskip
\item{37.}J.D. Bjorken, {\it Phys. Rev.} {\bf 148} (1966) 1467; {\it Phys. 
Rev.} {\bf D1} (1970) 1376.
\smallskip
\item{38.}J. Ellis and M. Karliner, {\it Phys. Lett.} {\bf B313} (1993) 
131, and CERN preprint TH.7022/93.
\smallskip
\item{39.}B. Adeva et al., {\it Phys. Lett.} {\bf B320} (1994) 400.
\bye